\documentclass[onecolumn,preprint,preprintnumbers,amsmath,amssymb,floatfix]{revtex4-1}

\usepackage{color}
\usepackage{graphicx}
\usepackage{float}
\usepackage{subfigure}

\begin{document}

\title{Shot noise variation within ensembles of gold atomic break junctions at room temperature}

\author{R.~Chen$^{1}$, M.~Matt$^{2}$, F.~Pauly$^{2}$, P.~Nielaba$^{2}$, J.~C.~Cuevas$^{3}$, D.~Natelson$^{1,4}$}
\email{natelson@rice.edu}
\thanks{corresponding author}

\affiliation{$^{1}$ Department of Physics and Astronomy, Rice University, 6100 Main St., Houston, 
TX 77005, USA}
\affiliation{$^{2}$ Department of Physics, University of Konstanz, D-78457 Konstanz, Germany}
\affiliation{$^{3}$ Departamento de F\'{\i}sica Te\'orica de la Materia Condensada and Condensed Matter Physics 
Center (IFIMAC), Universidad Aut\'onoma de Madrid, E-28049 Madrid, Spain}
\affiliation{$^{4}$ Department of Electrical and Computer Engineering, Rice University, 6100 Main St.,
Houston, TX 77005, USA}

\begin{abstract}
Atomic-scale junctions are a powerful tool to study quantum transport,
and are frequently examined through the mechanically controllable
break junction technique (MCBJ). The junction-to-junction variation of
atomic configurations often leads to a statistical approach, with
ensemble-averaged properties providing access to the relevant
physics. However, the full ensemble contains considerable additional
information.  We report a new analysis of shot noise over entire
ensembles of junction configurations using scanning tunneling
microscope (STM)-style gold break junctions at room temperature in
ambient conditions, and compare this data with simulations based on
molecular dynamics (MD), a sophisticated tight-binding model, and nonequilibrium
Green's functions. The experimental data show a suppression in the
variation of the noise near conductances dominated by fully
transmitting channels, and a surprising participation of multiple
channels in the nominal tunneling regime.  Comparison with the
simulations, which agree well with published work at low temperatures
and ultrahigh vacuum (UHV) conditions, suggests that these effects
likely result from surface contamination and disorder in the
electrodes.  We propose additional experiments that can distinguish
the relative contributions of these factors.
\end{abstract}

\maketitle

\section{Introduction}

Out of both fundamental and technological motivation, atomic-scale
junctions between conductors have been studied extensively in recent 
years \cite{Agrait:2003}. The size of the ``active'' region of such
devices is smaller than the inelastic mean free path for
electrons, and only a few quantum channels contribute to conduction;
thus such junctions serve as a simplified platform to study quantum
transport.  By inserting molecules between such electrodes, similar
physics studies have been extended to molecular junctions.  The
MCBJ technique enables study of
these junctions as a function of electrode separation, with each
junction configuration sampling from an ensemble of atomic
arrangements.  The varying mechanical and electrical properties of
junctions can be examined, enabling studies of conductance
quantization \cite{Muller:1996,Scheer:1997}, energy dissipation in
atomic ballistic wires \cite{Agrait:2002}, electron-phonon interaction
modifications to the shot noise \cite{Kumar:2012,Avriller:2012}, flicker noise in
metal junctions \cite{Wu:2008}, Joule heating in the electrodes 
\cite{Lee:2013}, and so on.

In most of these works, statistical analyses are usually conducted to
average out the microscopic variability of junction configurations.
For instance, conductance histograms
\cite{Itakura:1999,Yanson:2005,Venkataraman:2006} are a
common tool to study the preferred conductance values, and hence
preferred sets of discrete quantum channel transmittances, averaged
over accessible atomic configurations.  The obvious shortcoming is
that information associated with each specific atomic arrangement is
lost in the averaging.

To go beyond this ensemble-averaged analysis, considerable efforts
have been made in recent years.  Measurements of multiple Andreev
reflection reveal the transmittance of each specific quantum channel
in individual superconductor junctions \cite{Scheer:1997,Scheer:1998}.  
From ensembles, the range of mathematically allowed transmittances may be
estimated \cite{Vardimon:2013}. In the mean time, the development of
new analysis techniques also enables more information to be extracted
from the mountain of data over all the ensembles.  These methods
include ``density plots'', where every experimental data point in the
ensemble is counted.  Examples are density plots of conductance
vs.\ elongation \cite{Quek:2009,Kamenetska:2009,Vazquez:2012},
current-voltage characteristics
\cite{Lortscher:2007,Guedon:2012,Darancet:2012}, and
mechanical stiffness vs.\ elongation \cite{Fournier:2011}.
Two-dimensional (2D) cross-correlation methods also take advantage of
entire ensembles, revealing detailed information about motifs of
junction formation \cite{Halbritter:2010,Makk:2012,Aradhya:2013}.

Over the same time period, there have also been many advances in computational
techniques, particularly the ability to combine MD with
quantum transport to describe the inherent interplay between
mechanical and electrical properties in these systems, which is crucial
to establish a direct comparison with the experimental results 
\cite{Dreher:2005,Pauly:2006,Makk:2011,Schirm:2013}.

In this work, we analyse conductance and shot noise measurements over
whole ensembles of STM-style gold break junctions at room temperature,
expanding upon prior ensemble-averaged treatments
\cite{Wheeler:2010,Chen:2012,Chen:2014}. When mapping out
two dimensional density plots of shot noise vs.\ conductance, we find
the standard deviation (over the ensemble) in the noise at each
conductance, $G$.  At conductances in the nominal tunneling regime ($G
< 1 G_0 \equiv 2e^{2}/h$), we find a nonzero variance, which indicates
that in our system several quantum channels frequently contribute to
transport in this regime.  If normalized by the average noise, the
resulting fractional variance in the noise as a function of
conductance is relatively featureless, while the nonzero fractional
variance in the tunneling region still survives.  Furthermore, the
standard deviation of the shot noise shows clear minima coincident
with the ensemble-averaged shot noise spectral density suppressions
and conductance peaks near integer multiples of $G_0$. A state-of-the-art
calculation combining MD, an accurate tight-binding model, 
and Green's function techniques to examine junction formation is compared to 
these experimental results.  As we discuss below, surface contamination and
disorder in the electrodes are likely responsible for many of the differences
between the experiment and the calculations.

Shot noise, firstly discussed by Schottky in the context of vacuum
diodes \cite{Schottky:1918}, originates from the discreteness of charge
carriers \cite{Blanter:2000}. This nonequilibrium noise only exists in
addition to the Johnson-Nyquist noise \cite{Johnson:1928,Nyquist:1928}
at a finite bias. For Poisson-distributed, uncorrelated electrons,
the shot noise spectral density (A$^{2}$/Hz) is $S_I=2eI$, where $e$
is the electronic charge and $I$ is the average current.  Correlations
between the electrons modify the noise, which is often written in
the form $S_I=2eI F$, where $F$ is the Fano factor.  Measurements of
$F$ can therefore provide insight into the interactions of the
electrons with each other \cite{dePicciotto:1997,Saminadayar:1997} and 
other degrees of freedom such as phonon modes \cite{Kumar:2012,Avriller:2012}.

In atomic-scale junctions the electronic transport is generally described well by 
the Landauer-B{\"u}ttiker picture \cite{Datta:1995,DiVentra:2008,Cuevas:2010},
where electron-electron interactions are neglected and electrons transport 
coherently. The conductance is then $G=G_0\sum_i{\tau_i}$, where $\tau_i$ 
represents the transmittance of each discrete quantum channel indexed by $i$. 
The associated shot noise at zero temperature satisfies the following form
\cite{Khlus:1987,Lesovik:1989,Landauer:1991,Martin:1992,Buttiker:1992}
\begin{equation}
S_{I}=2eVG_{0}\sum_{i=1}^N{\tau_i(1-\tau_i)} ,
\label{eq:zeroT}
\end{equation}
where $V$ is the bias voltage across the junction. The corresponding Fano factor is
given by $F\equiv \sum_{i=1}^N{\tau_i(1-\tau_i)}/\sum_{i=1}^N{\tau_i}$, where $N$
is the number of open conduction channels. At elevated temperatures shot noise 
is enhanced and becomes entwined with Johnson-Nyquist noise. The total spectral 
density of the current noise is then
\begin{equation}
S_{I}=G_0 \left[ 4k_{\mathrm{B}}T\sum_{i=1}^N{\tau_i^{2}} + 2eV \coth 
\left(\frac{eV}{2k_{\mathrm{B}}T}\right) \sum_{i=1}^N{\tau_i(1-\tau_i)} \right] .
\label{eq:finiteT}
\end{equation}
This relation is relatively more complicated, but can still be
expressed in the form of the transmittance set \{$\tau_i$\}.  
We define the excess noise $P \equiv S_I(V)-S_I(0)$ as the
finite-temperature shot noise.  In a practical measurement, other
noise sources, such as flicker noise, can also contribute to the
measured noise, and must be considered case by case.  When shot noise
dominates the excess noise, it provides information about each quantum
channel through the $\tau_i(1-\tau_i)$ term, while the conductance
only reflects the overall contribution from all the channels.  In an
individual junction, if no more than two channels dominate conduction,
shot noise and conductance measurements together allow determination
of all the transport details. If more than two channels are involved,
the exact transmittance values cannot be completely determined, but
instead a mathematically allowed range of each transmittance may be
extracted \cite{Vardimon:2013}.

\section{Experimental methods}

The measurements are conducted across STM-style gold break junctions
at room temperature in air. As described in our previous
publications \cite{Chen:2012,Chen:2014}, a gold tip attached to an end
of one piezo actuator is electrically controlled, moving towards and
away from an evaporated thick gold film, allowing the cyclic formation
and breaking of gold junctions.  Only the data collected during the
breaking part of the cycle are analysed, with each breaking half cycle
defined as one trace.  At each fixed ``DC'' bias $V$, broad band (250
- 500~MHz) excess noise and DC conductance measurements are performed
simultaneously using a lock-in technique.  At the radio frequency
bandwidth, the dominant contributor of excess noise is shot noise, as
discussed in detail elsewhere \cite{Chen:2014}. The gold tip is
controlled to obtain approximately one trace per second.  Conductance
and noise power from lock-in outputs are acquired at $10^5$ samples/s,
but with the time constant on each lock-in set to 100~$\mu$s.  In each
trace, the time-averaged (with sub-millisecond averaging periods) shot
noise power as a function of conductance is computed.  Both noise
power and conductance axes are binned. The junction conductance is 
calculated from the measured current and applied voltage bias, taking 
into account series resistances in the measurement circuit. The 
statistical uncertainty in the conductance data is less than 1\%. 
The statistical uncertainty in the measurement of the rf power is at 
a similar level, though the background contributes a systematic 
uncertainty to individual measurements. The number of counts in each
bin $(G,P)$ is used to construct a 2D trace density plot. 

Typical traces have from five to more than ten thousand counts between 
$G = 4~G_{0}$ and $G=0.01~G_{0}$.  Traces with less than 1000 counts are not
included as these indicate anomalously rapid breakage or a measurement
problem; varying this cutoff from 500 counts to 2000 counts produces
no noticeable change in the resulting plots or analysis.  For clarity
of color scale when plotting the trace density, at each conductance
the bins are normalized to show the relative probability (between 0
and 1) of measuring a particular noise power.

In the analysis, a subtle technical issue arises regarding the
rigorous extraction of the shot noise from raw data.  The noise power
detector has small random fluctuations in addition to the true target
signal, causing a small but non-negligible positive background in the
lock-in amplifier measurement of its amplitude.  When performing a
full ensemble average analysis, statistical methods \cite{Chen:2012}
can be applied to remove this small background, as we have done
previously.  However, this background removal method does not work
rigorously at the single trace level; thus, in this work this
background remains.  Under this circumstance, there is therefore a
slight systematic overestimate of the true shot noise spectral
density, especially relevant when shot noise is small. The background 
is independent of $G$ and is equivalent to approximately $0.1
\times 10^{-24}$~A$^{2}$/Hz, and sets the ``floor'' of the data
shown in Fig.~1, panels (c) and (d).

\section{Experimental Results}

\begin{figure}[htb]                                                                     
\includegraphics[width=0.9\textwidth]{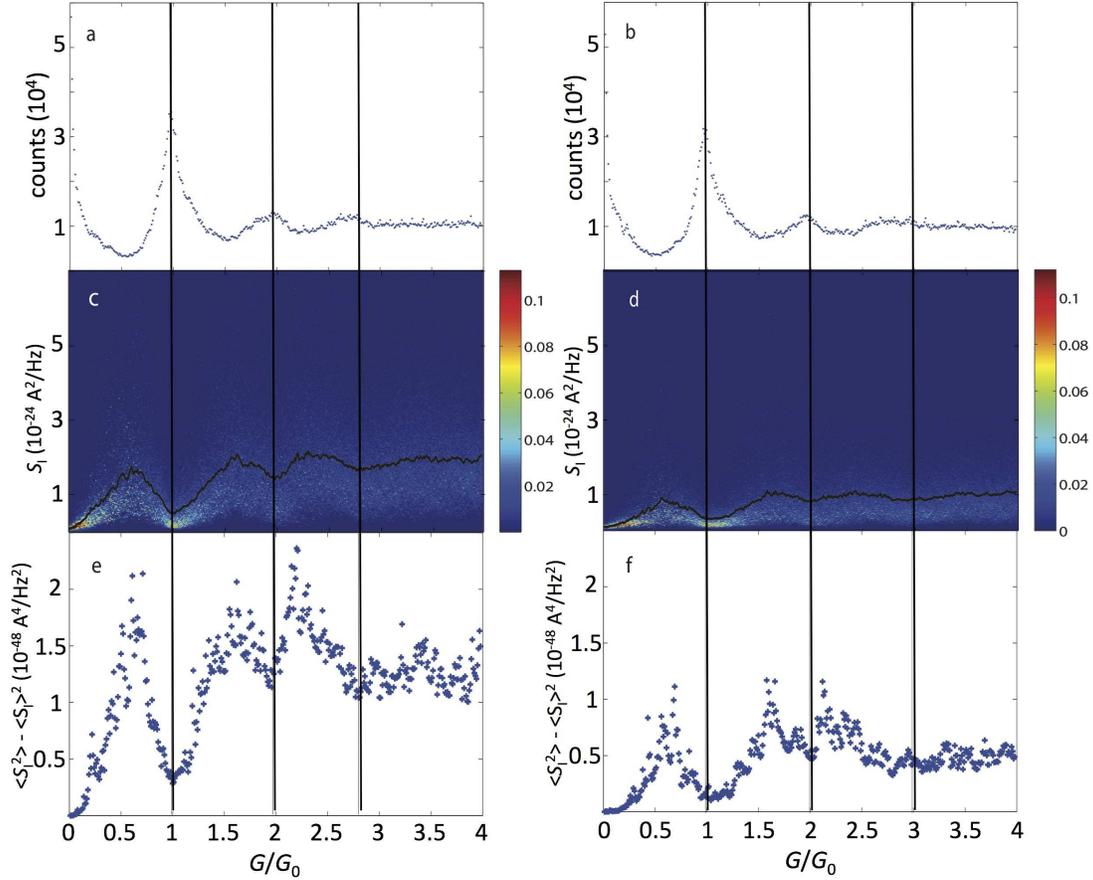}
\caption{Noise and its variation across ensembles.  (a) Conductance histogram 
acquired with 180 mV bias (816 traces). (b) Conductance histogram
acquired with 120 mV bias (807 traces). 
(c, d) 2D density plots for the respective ensembles, with data at each 
conductance normalized to show the probability for finding a particular shot 
noise power value at any particular conductance. The black curves show the 
ensemble-averaged noise power. (e, f) The associated variance of shot noise 
at each conductance. The vertical black lines are guides to the eye to indicate 
the conductance for which the shot noise variance is a
minimum.}
\label{Fig.1}
\end{figure}

Figure 1 shows the resulting density plots obtained at biases of
180~mV and 120~mV, respectively. The top-most panels are the
conductance histograms, commonly used to find the preferred
conductance values.  The middle panels are the trace density plots,
where warmer colors indicate higher densities of traces. The black
curve indicates the ensemble-averaged excess noise power as a function
of conductance.  From this panel, clearly the distribution of shot
noise has some interesting features.  The ``envelope'' of the traces
is comparatively large away from the conductances where shot noise is
suppressed, while that envelope becomes significantly concentrated at
those suppression regions near integer multiples of $G_0$.  

To quantify this distribution, the variance $\langle S_I^2 \rangle -
\langle S_I \rangle^2$ at each
conductance is computed and shown in panels (e, f). Note that the shot
noise variance minima are coincident with the conductance peaks and
noise suppressions.  Ensembles acquired at many other biases, or with
a different radio frequency bandwidth, reproduce the same features.
According to equations (1) and (2), shot noise suppressions originate from
the fully transmitting channels. It is also widely known that shot
noise suppressions in gold junctions coincide with the conductance
peaks \cite{Ludoph:2000}, indicating a potential relation between the
fully transmitted channels and one or a few preferred atomic
arrangements existing at the conductance plateaus.  There is no
obvious theoretical explanation for these shot noise variance minima.
We infer that somehow \{$\tau_i$\} has limited variations at these
preferred conductance values compared to higher or lower conductances.
Either relatively fewer atomic configurations are allowed at the
conductance plateaus, or the allowed atomic configurations have
relatively similar transmittance sets $\{\tau_i\}$, or both.

The density plots also reveal information about the nature of the shot 
noise suppression at the conductances indicated by the peaks in the 
conductance histograms. At cryogenic temperatures, theory and experiments 
agree very well that shot noise in gold junctions is suppressed down to 
near zero at conductance plateaus \cite{Ludoph:1999,vandenBrom:1999},
particularly when $G \approx G_{0}$. This indicates a single fully 
transmitted channel at $1G_0$ and some limited channel mixture at the 
higher $G$ plateaus. In our ensemble-averaged measurements, all the 
suppressions are only partial \cite{Chen:2012}. The density plot reveals 
the explanation: At $1G_0$, \textit{most} traces do show nearly complete 
suppression, while a small fraction relatively elevated noise. At other 
plateaus, though the lower count numbers and sparse distributions make 
the situation less obvious, the most populated shot noise values are also 
lower than the ensemble average. These observations show that the channel 
mixing responsible for affecting the depth of the noise suppression takes 
place only in a subset of the full ensemble. This is consistent with the 
channel mixing resulting from either particular realizations of disorder 
in the electrodes or surface contamination. The peaks in the conductance 
histograms are consistent with deviations from integer quantization in Au 
junctions previously ascribed to work hardening (and therefore disorder 
in the electrodes) \cite{Yanson:2005}.

The nonzero shot noise variance in the nominal tunneling regime also indicates 
a break-down of the single-channel picture below $1G_0$ in these ambient 
condition experiments. In single channel transport, the summation in equations (1) 
or (2) reduces to $\tau(1-\tau)$ while total conductance becomes proportional 
to $\tau$ of the only channel. The resulting Fano factor is $1-\tau$, and at 
a given $G$ its variance is zero. Thus a nonzero noise variance when $G < G_{0}$ 
clearly indicates the participation of multiple quantum channels. The 
diffusive-like features at high conductance reported in our former paper also 
indicate that the channel mixture in the high conductance region in our measurements 
is stronger than commonly expected \cite{Chen:2012}.

\begin{figure}[htb]                                                                     
\includegraphics[width=0.5\textwidth]{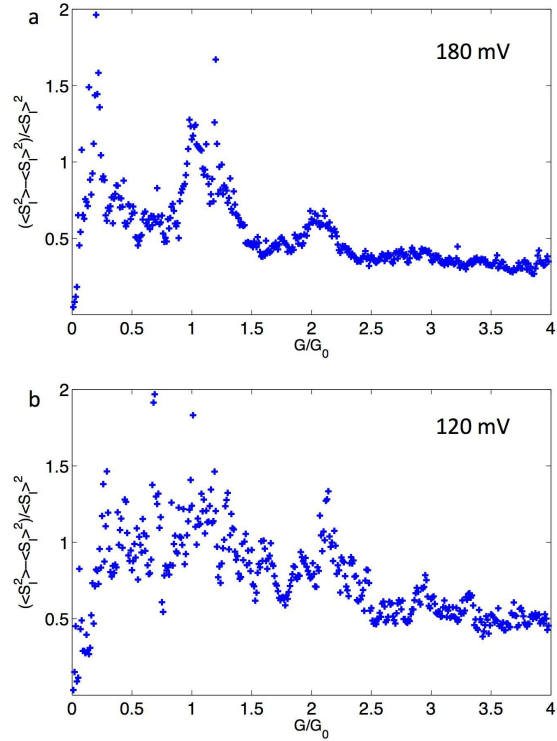}
\caption{Fractional variance of the noise power as a function of conductance, 
using the same data as in figure 1.}
\label{Fig.2}
\end{figure}

A debatable question here is whether the variance $\langle S_I^2 \rangle - 
\langle S_I \rangle^2$ is a fair way to examine shot noise's variation over 
ensembles. In figure~\ref{Fig.2} we plot the fractional variance $(\langle S_I^2 \rangle
- \langle S_I \rangle^2)/ \langle S_I \rangle^2$, using the same data as in 
figure 1. After this normalization, the fractional variance as a function of 
conductance tends to be relatively featureless or even have maxima at the 
preferred conductances. This suggests that the variance in the noise is 
approximately proportional to the square of the noise itself, though there is 
no obvious explanation for this. The nonzero noise variance below $1G_0$ 
naturally remains.

\section{Simulation methods}

This section is devoted to the description of the theoretical methods that we
employed to shed some light on our experimental results for the shot noise
of Au atomic-size contacts. For this purpose, we combined classical MD
simulations of the formation of atomic-size contacts with a 
tight-binding description of the electronic structure to compute the transport 
properties with the help of Green's function techniques. Our approach follows
closely Refs. \cite{Dreher:2005,Pauly:2006,Pauly:2011,Schirm:2013}, and
we now proceed to explain it in some detail.

\begin{figure}[t]
\includegraphics[width=0.8\textwidth]{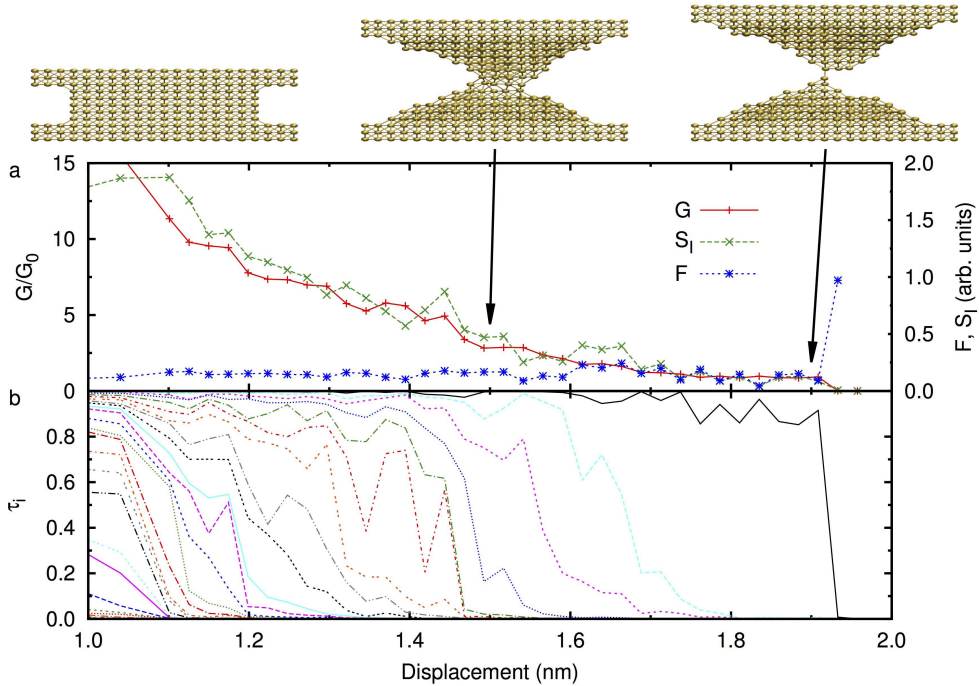}
\caption{Example of our simulations of the stretching of a Au wire at 300 K.
(a) Conductance, shot noise, and Fano factor as a function of the elongation of
the wire. (b) Individual transmission coefficients as a function of the elongation.
The upper panels show the wire geometries at different stages of the elongation
process. The left geometry corresponds to the initial configuration of the
simulated wire.}
\label{thy-example}
\end{figure}

In these atomic wires there is an intricate relation between the
mechanical and the transport properties. For this reason, and in order
to establish a direct comparison with our experiments, it is necessary
to describe the formation process of these atomic-size contacts. For
this purpose, we carried out classical MD simulations using the open
source program package LAMMPS \cite{Plimpton:1995,LAMMPS}.  These
simulations are based on the so-called embedded atom method and, in
particular, we employed the empirical potentials from Ref.~\cite{Sheng:2011}. 
Let us emphasize that these potentials overcome several 
problems of two-body potentials like, for instance, the coordination 
independence of the bond strength. This is important for our calculations
because in our atomic contacts we have, in particular, regions with
low coordination number. To generate the geometrical configurations,
we start with an ideal face-centered cubic lattice, where the crystal direction
$\langle 100 \rangle$ lies parallel to the $z$ axis, which corresponds
to the transport and elongation direction.  For the MD calculations, we
divided the wire geometry into three parts:  Two electrodes connected by a
central wire (see figure \ref{thy-example}). The
electrodes consist of 661 atoms each and they are kept fixed during
the MD calculations. On the other hand, the wire is made up of 563
atoms which follow the Newtonian equations of motion.  We assume a
canonical ensemble and use the velocity Verlet integration scheme 
\cite{Frenkel:2004}. The
wire has an initial length of 0.83 nm and the starting velocities of
the atoms in the wire were chosen randomly with a Gaussian
distribution to yield the desired average temperature. As we discuss
below, we performed simulations at room temperature ($T = 300$~K), but
also at cryogenic temperatures ($T = 4$~K) to compare with previously
published results. Because of the randomness in the initial
velocities, every elongation process evolves differently, while a
Nose-Hoover thermostat keeps the temperature fixed \cite{Frenkel:2004}. 
To relax the system, the wire was equilibrated for 0.1~ns. Finally, the 
elongation process is simulated by separating the electrodes at a constant
velocity of 0.4 m/s. During this process, every 10 ps the geometry is
recorded. A stretching process needs a total simulation time of about
4.5 ns, until the contact breaks.

Once the geometries of the atomic wires were determined through the MD
simulations, we used them to compute the conductance and the shot
noise in the spirit of the Landauer-B\"uttiker formalism. As explained
in the introduction, within this formalism the transport properties
are fully determined by the set of transmission coefficients
$\{\tau_i\}$. We computed these coefficients by combining an
appropriate description of the electronic structure of the Au wires
with non-equilibrium Green's function techniques following a standard
recipe that we have explained in detail elsewhere
\cite{Dreher:2005,Pauly:2008,Cuevas:2010}.  The
electronic structure was described within the framework of the
Slater-Koster tight-binding parameterization of Ref.~\cite{Mehl:1998},
which has been quite successful in determining a variety of properties of
these atomic wires \cite{Cuevas:2010}.  To compute the transmission
coefficients with the help of the Green's function techniques, we
divided the system into three regions as in the MD
simulations, \emph{i.e.}\ the upper and lower electrodes and the
central wire. The electrodes were considered to be semi-infinite
perfect crystals. Their surface Green's functions were computed with
the help of a decimation technique \cite{Guinea:1983,Pauly:2008},
using the same tight-binding parameterization as for the
central part to determine their electronic structure. It is worth
stressing that we enforced charge neutrality for all the atoms of the
wire, which is a condition that is typically fulfilled in metallic
systems \cite{Brandbyge:1999}.

Finally, to accumulate sufficient statistics for our study, we carried 
out 100 MD simulations of the breaking of the Au atomic wires and 
checked that this number is sufficient, especially, to converge the 
conductance histograms. We show in figure \ref{thy-example} a typical 
example of the simulation of stretching of a wire at room temperature. 
In particular, we show the different physical properties of interest 
in the last stages of the breaking of the wire, namely the conductance, 
shot noise, and Fano factor. Additionally, we show the individual 
transmission coefficients $\tau_i$ as a function of the elongation.

\section{Simulation results and comparison with the experiments}

Let us now discuss the main results of the simulations described in
the previous section.  We stress that we shall be
discussing results at room temperature ($T=300$ K) to compare directly
with the experimental results presented above, but we shall also
present results at low temperatures ($T=4$ K) to compare with
previously published experimental results obtained at cryogenic
temperatures and under ultra-high vacuum conditions
\cite{vandenBrom:1999,Vardimon:2013}.

\begin{figure}[t]
\includegraphics[width=0.9\textwidth]{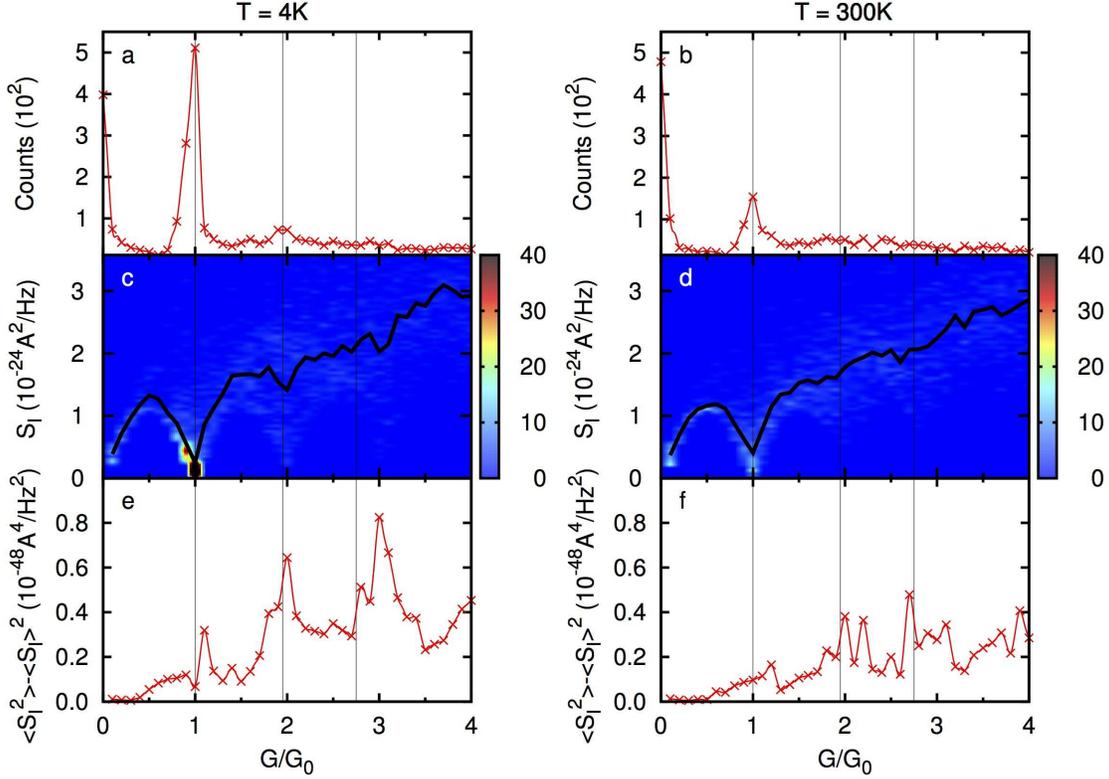}
\caption{(a,b) Conductance histogram obtained from 100 MD simulations
at 4 and 300 K, respectively. (c,d) The corresponding 2D density
plots for the shot noise.  The black solid lines indicate the
ensemble-averaged values. (e,f) The associated variance of the shot
noise as a function of the conductance. The bias voltage for the
calculation of $S_I$ is set to 180 mV. The vertical lines are at
the same position as in figure~\ref{Fig.1}.}
\label{thy-noise}
\end{figure}

Our main results are summarized in figure~\ref{thy-noise}, where we
show the conductance histograms, density plots of the shot noise, and
the noise variance as a function of the conductance for 4 and 300 K.
The conductance and the noise were computed in the linear response
regime, \emph{i.e.}\ using the zero-bias transmission coefficients. 
Moreover, for the noise, we focus on the shot noise contribution to the
excess noise by merely plotting the quantity $2eG_0 V 
\sum_{i=1}^N{\tau_i(1-\tau_i)}$ in equation (2), which ignores the
contribution of the thermal noise. This is justified as $k_{\rm B}T$
is much smaller than $eV$ in our experiments. For comparison with the
experiments we have set the bias voltage in the prefactor to 180 mV. As one 
can see in panels (a) and (b), the conductance histograms are dominated 
by a pronounced peak at $1G_0$, which is much higher at low
temperatures.   This is due to the fact that this peak partially
originates from the formation of monoatomic chains
\cite{Ohnishi:1998,Yanson:1998}.  These chains exhibit a conductance
very close to $1G_0$ and we find that their formation is clearly less
favorable at room temperature.  Notice also the appearance of another
pronounced peak at $2G_0$ at low temperatures, which is to a large
extent washed out at room temperature.  Turning to the noise 
(figure~\ref{thy-noise}(c,d)), the most noticeable feature is its
suppression close to multiples of $G_0$, which is specially pronounced
around $1G_0$ and at $T=4$ K. This is consistent with the measurements
reported above and supports the assumption that the channel
mixture is stronger outside the tunneling region at higher
temperatures.  On the other hand, the variance of the shot noise
(figure~\ref{thy-noise}(e,f)) generally exhibits \textit{maxima} correlated 
with the maxima in the conductance histograms. While this is consistent with
the low-temperature results reported by Vardimon \emph{et
  al.}\ \cite{Vardimon:2013a}, this is clearly at odds with our
experimental results (figure~\ref{Fig.1}(e,f)). Notice also that
the variance of the noise does not vanish below $1G_0$, which clearly
suggests that several channels are contributing in that region.

\begin{figure}[t]
\includegraphics[width=0.9\textwidth]{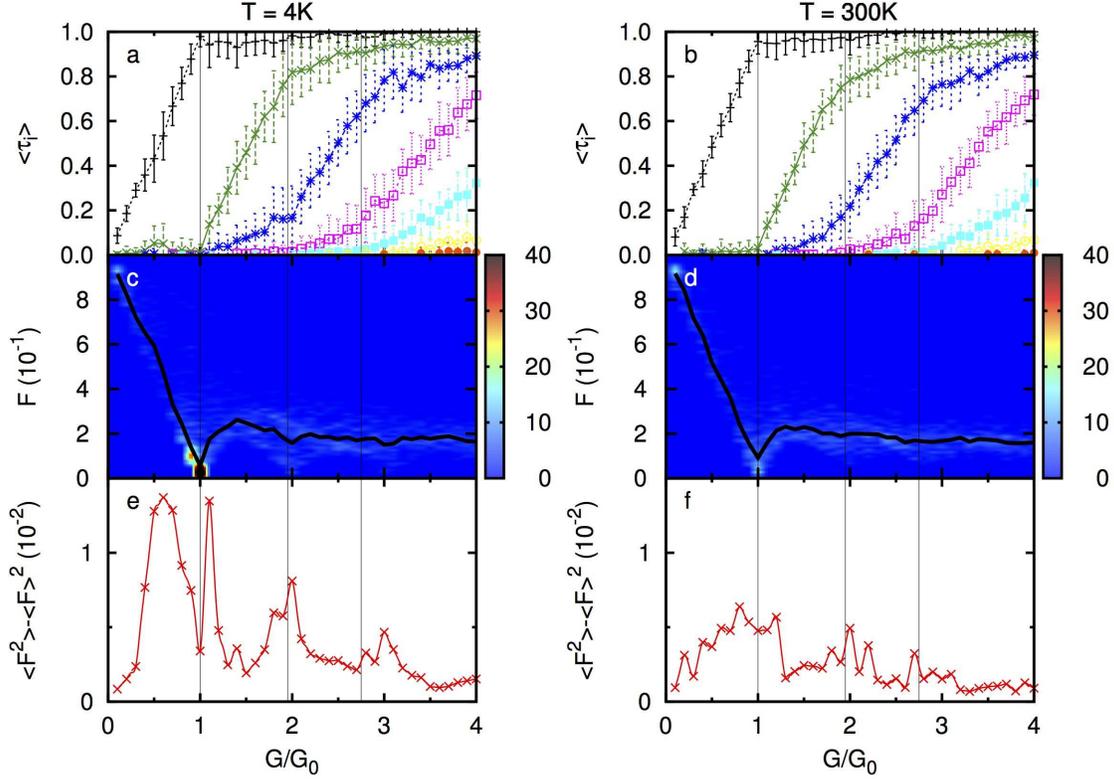}
\caption{(a,b) Channel distributions corresponding to the results of
  figure \ref{thy-noise} at temperatures of 4 and 300 K, respectively. (c,d) The
  corresponding 2D density plots for the Fano factor.  The black solid
  lines indicate the ensemble-averaged values. (e,f) The associated
  variance of the Fano factor as a function of the conductance.}
\label{thy-fano}
\end{figure}

The origin of the results just described can be clarified by analysing
both the distribution of transmission coefficients, $\{\tau_i\}$, and
the Fano factor and its corresponding variance. This information is
displayed in figure~\ref{thy-fano}.  The channel distributions shown
in panels (a) and (b) indicate that the conduction channels open one
by one as the conductance increases, as reported in
Ref.~\cite{Scheer:1998} with the help of proximity-induced
superconductivity in Au atomic-size contacts.  This is a typical
behavior of a monovalent metal.  Focusing on the low temperature
results, we find that although the conductance region $G < G_0$ is
largely dominated by a single channel, a second one gives a
sizeable contribution.  This additional channel manifests as a kink in
the Fano factor around $0.5G_0$ and as a pronounced maximum in the
Fano factor variance at that position.  It is worth stressing that the
presence of this second channel does not ``spoil" the conductance
quantization and the histogram still exhibits a very pronounced peak
at $1G_0$.  For higher conductance, the Fano factor exhibits a partial
suppression close to multiples of $G_0$, while its variance has maxima
correlated with the maxima of the conductance or the minima of the
shot noise.  All these findings at 4 K are in very good agreement with
the recent experimental results of Refs.~\cite{Vardimon:2013,Vardimon:2013a} 
where shot noise was measured at cryogenic temperatures and used to
extract the channel distribution of Au few-atom contacts. Turning now
to the room temperature results, one can still see a small contribution 
of a second channel for $G < G_0$, which explains the non-vanishing 
variance of both the shot noise and the Fano factor for these conductances. 
However, at 300 K the weight of this channel is not sufficient to produce 
a maximum in the variance of these quantities below $G_0$. For higher 
conductance, the variance of the Fano factor exhibits again shallow 
maxima correlated with the maxima of the conductance and the minima of 
the shot noise.

\begin{figure}[t]
\includegraphics[width=0.7\textwidth]{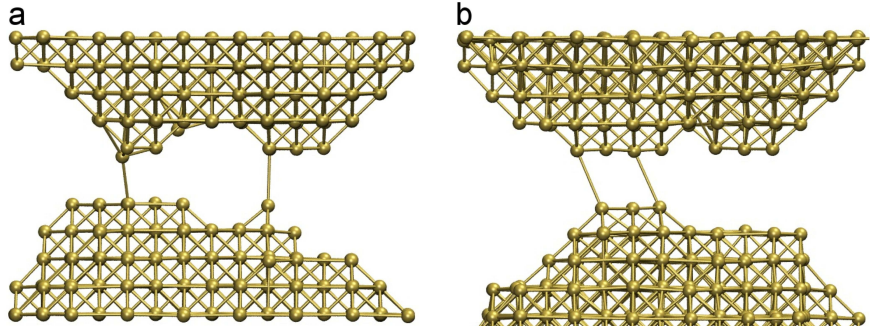}
\caption{Two representative examples of contact geometries where two conduction channels 
give a significant contribution in the tunnel regime ($G< 0.5G_0$). The total 
conductance and the individual transmission coefficients in the different examples 
are (where we consider only channels with $\tau_i>0.01$): (a) $G = 0.29G_0$, 
$\tau_1 =0.28$, $\tau_2 = 0.11$, $\tau_3 = 0.01 $ and (b) $G = 0.15G_0$, 
$\tau_1 =0.09$, $\tau_2 = 0.05$, and $\tau_3 = 0.01$.} 
\label{thy-geo-tunnel}
\end{figure}

From the previous discussion, a natural question arises: What is the 
origin of the second channel that appears in some geometries in the 
tunnel regime? To answer it we analysed systematically those contact 
geometries where at least two channels contribute significantly
to the transport properties in the conductance region $G < 0.5 G_0$.
Some representative examples are shown in figure~\ref{thy-geo-tunnel}.
We found that these geometries can be grouped into two categories. In the 
first one, the electrodes contain several neighbouring atoms that are at
similar distances to the atoms in the other electrode, see 
figure~\ref{thy-geo-tunnel}(b). At room temperature, practically all
the geometries with two channels in the tunnel regime belong to this kind.
On the other hand, at 4 K we also find geometries where the contact breaks
at several points giving rise to parallel junctions, see figure~\ref{thy-geo-tunnel}(a).
In both types of geometries the two channels originate from parallel paths
with the peculiarity that in the second kind these paths are basically
independent, \emph{i.e.}\ they do not interfere. Interestingly, we find
that the geometries with several parallel junctions are also responsible 
for the peak at $2G_0$ that appears in the low temperature conductance
histogram.

In summary, our low temperature theoretical results are in very
good agreement with cryogenic experiments performed in clean
conditions and our room temperature results supports the idea of
several channels contributing below $1G_0$. However, our theoretical data are
not able to reproduce our room temperature experimental findings
related to the presence of minima in the variance of the shot noise
and Fano factor correlated with the minima in their averaged values.
The peaks in the experimental conductance histograms,
particularly those above $1~G_{0}$, are also considerably more robust than
what is seen in the simulations. The survival of higher quantization peaks
at room temperature has been observed in many experiments 
\cite{Yanson:2005,Gai:1996,Muller:1996,CostaKramer:1997}.

\section{Further discussion and conclusions}

As discussed in the previous section, our theoretical simulations
have difficulties to reproduce the observed shot noise variance minima, 
and the robust quantization peaks and noise minima at $G > G_{0}$ at room 
temperature. The origins of these discrepancies are not clear at the moment. 
One possibility might be related to the unavoidable limitations in our MD simulations. 
In particular, one has to bear in mind that the stretching velocity in
these simulations is many orders of magnitudes higher than in the
experiments. Moreover, because of computational reasons we have to
constrain ourselves to relatively small contacts.  Although we have
investigated these issues systematically within our capabilities and
have not found any significant difference in our results upon
changes in the stretching velocity, we cannot rule out that
this discrepancy is due to systematic deficiencies in our simulations.
However, it is worth remarking that our low temperature
simulations are in very good agreement with published data in UHV
conditions \cite{Vardimon:2013} and similar simulations have been very
successful in reproducing, for instance, the thermopower of different
atomic wires \cite{Pauly:2011} or the channel distribution of Al
atomic-size contacts at very low temperatures \cite{Schirm:2013}.

Another possibility for this discrepancy might be the fact that these 
experiments are conducted in air. Though the related environmental
parameters such as pressure, temperature or humidity should be
relatively stable at each set of measurements, contamination from adsorbates 
(water molecules for instance) cannot be ruled out.  Molecules randomly
attached to the junction can surely alter the dynamics and stability of the
junctions and the conductance as well, especially in the last stages
of the breaking process. Both hydrogen \cite{Csonka:2003,Barnett:2004} and
oxygen \cite{Thijssen:2008} are known to incorporate into junctions and 
alter the available stable atomic configurations. This possibility can be 
tested by further room temperature experiments in a UHV environment. On the 
other hand, the disorder of gold junctions may also be relevant \cite{Yanson:2005}, 
which is proven to have effect on peak positions in conductance histogram. Dislocations 
and other defects introduced by ``work hardening'' can in principle affect atomic 
mobility and therefore structural stability, physics very difficult to capture 
in computationally tractable simulations.

To summarize, we have reported measurements of the shot noise's
distribution over ensembles on STM-style gold break junctions at room
temperature. The details of the partial noise suppressions seen in the 
ensemble average results are revealed, showing that outliers can obscure 
a subensemble with much greater noise suppression compatible with fully 
transmitting channels. We find signs of non-negligible conduction from 
additional quantum channels in the nominal tunneling region, and observe 
minima in the shot noise variance coincident with conductance histogram 
peaks and averaged shot noise suppressions. Simulations combining MD, a
tight-binding model and Green's functions techniques have been
conducted. The simulations successfully reproduce many key features of the
data and confirm a second channel's participation below $1G_0$. Other aspects
of the data, such as the minima in the noise variance near the preferred
conductance values and robust conductance histogram peaks above $1G_0$, are
not seen in the simulations. One possible explanation is that the simulations 
are restricted to comparatively small contact geometries and might therefore 
miss some junction configurations possible in the experiments. Another factor 
responsible for these discrepancies may be surface contamination in the experiments, 
altering the breaking dynamics, an effect that is not taken into account in 
the simulations. Adsorbed contaminants can hinder diffusion and alter the 
breaking dynamics, effects not simulated here. Structural disorder, leading to 
modifications in the preferred conductances as well as potentially hindering 
atomic rearrangements, may also be relevant. Further experiments in UHV and
under various electrode annealing conditions as well as simulations with yet
larger atomic contacts may be able to resolve the relative contributions of 
these candidate mechanisms.

\section{Acknowledgments}

D.N.\ and R.C.\ acknowledge support of NSF awards DMR-0855607 and
DMR-1305879, as well as helpful discussions with P. J. Wheeler. 
F.P.\ acknowledges the funding from the Carl Zeiss foundation and the 
collaborative research center SFB 767. J.C.C.\ acknowledges financial 
support from the Spanish MICINN (Contract No.\ FIS2011-28851-C02-01) and 
thanks Ran Vardimon, Oren Tal, and Nicol\'as Agra{\"i}t for useful 
discussions.  The authors thank the NIC for computer time.

\clearpage


\begin{thebibliography}{64}
\expandafter\ifx\csname natexlab\endcsname\relax\def\natexlab#1{#1}\fi
\expandafter\ifx\csname bibnamefont\endcsname\relax
  \def\bibnamefont#1{#1}\fi
\expandafter\ifx\csname bibfnamefont\endcsname\relax
  \def\bibfnamefont#1{#1}\fi
\expandafter\ifx\csname citenamefont\endcsname\relax
  \def\citenamefont#1{#1}\fi
\expandafter\ifx\csname url\endcsname\relax
  \def\url#1{\texttt{#1}}\fi
\expandafter\ifx\csname urlprefix\endcsname\relax\def\urlprefix{URL }\fi
\providecommand{\bibinfo}[2]{#2}
\providecommand{\eprint}[2][]{\url{#2}}

\bibitem[{\citenamefont{Agra{\"i}t et~al.}(2003)\citenamefont{Agra{\"i}t,
  Yeyati, and van Ruitenbeek}}]{Agrait:2003}
\bibinfo{author}{\bibfnamefont{N.}~\bibnamefont{Agra{\"i}t}},
  \bibinfo{author}{\bibfnamefont{A.~L.} \bibnamefont{Yeyati}},
  \bibnamefont{and} \bibinfo{author}{\bibfnamefont{J.~M.} \bibnamefont{van
  Ruitenbeek}}, \bibinfo{journal}{Phys. Rep.} \textbf{\bibinfo{volume}{377}},
  \bibinfo{pages}{81 } (\bibinfo{year}{2003}).

\bibitem[{\citenamefont{Muller et~al.}(1996)\citenamefont{Muller, Krans,
  Todorov, and Reed}}]{Muller:1996}
\bibinfo{author}{\bibfnamefont{C.~J.} \bibnamefont{Muller}},
  \bibinfo{author}{\bibfnamefont{J.~M.} \bibnamefont{Krans}},
  \bibinfo{author}{\bibfnamefont{T.~N.} \bibnamefont{Todorov}},
  \bibnamefont{and} \bibinfo{author}{\bibfnamefont{M.~A.} \bibnamefont{Reed}},
  \bibinfo{journal}{Phys. Rev. B} \textbf{\bibinfo{volume}{53}},
  \bibinfo{pages}{1022} (\bibinfo{year}{1996}).

\bibitem[{\citenamefont{Scheer et~al.}(1997)\citenamefont{Scheer, Joyez,
  Esteve, Urbina, and Devoret}}]{Scheer:1997}
\bibinfo{author}{\bibfnamefont{E.}~\bibnamefont{Scheer}},
  \bibinfo{author}{\bibfnamefont{P.}~\bibnamefont{Joyez}},
  \bibinfo{author}{\bibfnamefont{D.}~\bibnamefont{Esteve}},
  \bibinfo{author}{\bibfnamefont{C.}~\bibnamefont{Urbina}}, \bibnamefont{and}
  \bibinfo{author}{\bibfnamefont{M.~H.} \bibnamefont{Devoret}},
  \bibinfo{journal}{Phys. Rev. Lett.} \textbf{\bibinfo{volume}{78}},
  \bibinfo{pages}{3535} (\bibinfo{year}{1997}).

\bibitem[{\citenamefont{Agra{\"i}t et~al.}(2002)\citenamefont{Agra{\"i}t,
  Untiedt, Rubio-Bollinger, and Vieira}}]{Agrait:2002}
\bibinfo{author}{\bibfnamefont{N.}~\bibnamefont{Agra{\"i}t}},
  \bibinfo{author}{\bibfnamefont{C.}~\bibnamefont{Untiedt}},
  \bibinfo{author}{\bibfnamefont{G.}~\bibnamefont{Rubio-Bollinger}},
  \bibnamefont{and} \bibinfo{author}{\bibfnamefont{S.}~\bibnamefont{Vieira}},
  \bibinfo{journal}{Phys. Rev. Lett.} \textbf{\bibinfo{volume}{88}},
  \bibinfo{pages}{216803} (\bibinfo{year}{2002}).

\bibitem[{\citenamefont{Kumar et~al.}(2012)\citenamefont{Kumar, Avriller,
  Yeyati, and van Ruitenbeek}}]{Kumar:2012}
\bibinfo{author}{\bibfnamefont{M.}~\bibnamefont{Kumar}},
  \bibinfo{author}{\bibfnamefont{R.}~\bibnamefont{Avriller}},
  \bibinfo{author}{\bibfnamefont{A.~L.} \bibnamefont{Yeyati}},
  \bibnamefont{and} \bibinfo{author}{\bibfnamefont{J.~M.} \bibnamefont{van
  Ruitenbeek}}, \bibinfo{journal}{Phys. Rev. Lett.}
  \textbf{\bibinfo{volume}{108}}, \bibinfo{pages}{146602}
  (\bibinfo{year}{2012}).

\bibitem[{\citenamefont{Avriller and Frederiksen}(2012)}]{Avriller:2012}
\bibinfo{author}{\bibfnamefont{R.}~\bibnamefont{Avriller}} \bibnamefont{and}
  \bibinfo{author}{\bibfnamefont{T.}~\bibnamefont{Frederiksen}},
  \bibinfo{journal}{Phys. Rev. B} \textbf{\bibinfo{volume}{86}},
  \bibinfo{pages}{155411} (\bibinfo{year}{2012}).

\bibitem[{\citenamefont{Wu et~al.}(2008)\citenamefont{Wu, Wu, Oberholzer,
  Steinacher, Calame, and Sch\"onenberger}}]{Wu:2008}
\bibinfo{author}{\bibfnamefont{Z.}~\bibnamefont{Wu}},
  \bibinfo{author}{\bibfnamefont{S.}~\bibnamefont{Wu}},
  \bibinfo{author}{\bibfnamefont{S.}~\bibnamefont{Oberholzer}},
  \bibinfo{author}{\bibfnamefont{M.}~\bibnamefont{Steinacher}},
  \bibinfo{author}{\bibfnamefont{M.}~\bibnamefont{Calame}}, \bibnamefont{and}
  \bibinfo{author}{\bibfnamefont{C.}~\bibnamefont{Sch\"onenberger}},
  \bibinfo{journal}{Phys. Rev. B} \textbf{\bibinfo{volume}{78}},
  \bibinfo{pages}{235421} (\bibinfo{year}{2008}).

\bibitem[{\citenamefont{Lee et~al.}(2013)\citenamefont{Lee, Kim, Jeong, Zotti,
  Pauly, Cuevas, and Reddy}}]{Lee:2013}
\bibinfo{author}{\bibfnamefont{W.}~\bibnamefont{Lee}},
  \bibinfo{author}{\bibfnamefont{K.}~\bibnamefont{Kim}},
  \bibinfo{author}{\bibfnamefont{W.}~\bibnamefont{Jeong}},
  \bibinfo{author}{\bibfnamefont{L.~A.} \bibnamefont{Zotti}},
  \bibinfo{author}{\bibfnamefont{F.}~\bibnamefont{Pauly}},
  \bibinfo{author}{\bibfnamefont{J.~C.} \bibnamefont{Cuevas}},
  \bibnamefont{and} \bibinfo{author}{\bibfnamefont{P.}~\bibnamefont{Reddy}},
  \bibinfo{journal}{Nature} \textbf{\bibinfo{volume}{498}},
  \bibinfo{pages}{209} (\bibinfo{year}{2013}).

\bibitem[{\citenamefont{Itakura et~al.}(1999)\citenamefont{Itakura, Yuki,
  Kurokawa, Yasuda, and Sakai}}]{Itakura:1999}
\bibinfo{author}{\bibfnamefont{K.}~\bibnamefont{Itakura}},
  \bibinfo{author}{\bibfnamefont{K.}~\bibnamefont{Yuki}},
  \bibinfo{author}{\bibfnamefont{S.}~\bibnamefont{Kurokawa}},
  \bibinfo{author}{\bibfnamefont{H.}~\bibnamefont{Yasuda}}, \bibnamefont{and}
  \bibinfo{author}{\bibfnamefont{A.}~\bibnamefont{Sakai}},
  \bibinfo{journal}{Phys. Rev. B} \textbf{\bibinfo{volume}{60}},
  \bibinfo{pages}{11163} (\bibinfo{year}{1999}).

\bibitem[{\citenamefont{Yanson et~al.}(2005)\citenamefont{Yanson,
  Shklyarevskii, Csonka, van Kempen, Speller, Yanson, and van
  Ruitenbeek}}]{Yanson:2005}
\bibinfo{author}{\bibfnamefont{I.~K.} \bibnamefont{Yanson}},
  \bibinfo{author}{\bibfnamefont{O.~I.} \bibnamefont{Shklyarevskii}},
  \bibinfo{author}{\bibfnamefont{S.}~\bibnamefont{Csonka}},
  \bibinfo{author}{\bibfnamefont{H.}~\bibnamefont{van Kempen}},
  \bibinfo{author}{\bibfnamefont{S.}~\bibnamefont{Speller}},
  \bibinfo{author}{\bibfnamefont{A.~I.} \bibnamefont{Yanson}},
  \bibnamefont{and} \bibinfo{author}{\bibfnamefont{J.~M.} \bibnamefont{van
  Ruitenbeek}}, \bibinfo{journal}{Phys. Rev. Lett.}
  \textbf{\bibinfo{volume}{95}}, \bibinfo{pages}{256806}
  (\bibinfo{year}{2005}).

\bibitem[{\citenamefont{Venkataraman et~al.}(2006)\citenamefont{Venkataraman,
  Klare, Tam, Nuckolls, Hybertsen, and Steigerwald}}]{Venkataraman:2006}
\bibinfo{author}{\bibfnamefont{L.}~\bibnamefont{Venkataraman}},
  \bibinfo{author}{\bibfnamefont{J.~E.} \bibnamefont{Klare}},
  \bibinfo{author}{\bibfnamefont{I.~W.} \bibnamefont{Tam}},
  \bibinfo{author}{\bibfnamefont{C.}~\bibnamefont{Nuckolls}},
  \bibinfo{author}{\bibfnamefont{M.~S.} \bibnamefont{Hybertsen}},
  \bibnamefont{and} \bibinfo{author}{\bibfnamefont{M.~L.}
  \bibnamefont{Steigerwald}}, \bibinfo{journal}{Nano Lett.}
  \textbf{\bibinfo{volume}{6}}, \bibinfo{pages}{458} (\bibinfo{year}{2006}).

\bibitem[{\citenamefont{Scheer et~al.}(1998)\citenamefont{Scheer, Agra{\"i}t,
  Cuevas, Yeyati, Ludoph, Mart{\'i}n-Rodero, Bollinger, van Ruitenbeek, and
  Urbina}}]{Scheer:1998}
\bibinfo{author}{\bibfnamefont{E.}~\bibnamefont{Scheer}},
  \bibinfo{author}{\bibfnamefont{N.}~\bibnamefont{Agra{\"i}t}},
  \bibinfo{author}{\bibfnamefont{J.~C.} \bibnamefont{Cuevas}},
  \bibinfo{author}{\bibfnamefont{A.~L.} \bibnamefont{Yeyati}},
  \bibinfo{author}{\bibfnamefont{B.}~\bibnamefont{Ludoph}},
  \bibinfo{author}{\bibfnamefont{A.}~\bibnamefont{Mart{\'i}n-Rodero}},
  \bibinfo{author}{\bibfnamefont{G.~R.} \bibnamefont{Bollinger}},
  \bibinfo{author}{\bibfnamefont{J.~M.} \bibnamefont{van Ruitenbeek}},
  \bibnamefont{and} \bibinfo{author}{\bibfnamefont{C.}~\bibnamefont{Urbina}},
  \bibinfo{journal}{Nature} \textbf{\bibinfo{volume}{394}},
  \bibinfo{pages}{154} (\bibinfo{year}{1998}).

\bibitem[{\citenamefont{Vardimon et~al.}(2013)\citenamefont{Vardimon, Klionsky,
  and Tal}}]{Vardimon:2013}
\bibinfo{author}{\bibfnamefont{R.}~\bibnamefont{Vardimon}},
  \bibinfo{author}{\bibfnamefont{M.}~\bibnamefont{Klionsky}}, \bibnamefont{and}
  \bibinfo{author}{\bibfnamefont{O.}~\bibnamefont{Tal}},
  \bibinfo{journal}{Phys. Rev. B} \textbf{\bibinfo{volume}{88}},
  \bibinfo{pages}{161404} (\bibinfo{year}{2013}).

\bibitem[{\citenamefont{Quek et~al.}(2009)\citenamefont{Quek, Kamenetska,
  Steigerwald, Choi, Louie, Hybertsen, Neaton, and Venkataraman}}]{Quek:2009}
\bibinfo{author}{\bibfnamefont{S.~Y.} \bibnamefont{Quek}},
  \bibinfo{author}{\bibfnamefont{M.}~\bibnamefont{Kamenetska}},
  \bibinfo{author}{\bibfnamefont{M.~L.} \bibnamefont{Steigerwald}},
  \bibinfo{author}{\bibfnamefont{H.~J.} \bibnamefont{Choi}},
  \bibinfo{author}{\bibfnamefont{S.~G.} \bibnamefont{Louie}},
  \bibinfo{author}{\bibfnamefont{M.~S.} \bibnamefont{Hybertsen}},
  \bibinfo{author}{\bibfnamefont{J.~B.} \bibnamefont{Neaton}},
  \bibnamefont{and}
  \bibinfo{author}{\bibfnamefont{L.}~\bibnamefont{Venkataraman}},
  \bibinfo{journal}{Nat. Nanotechnol.} \textbf{\bibinfo{volume}{4}},
  \bibinfo{pages}{230} (\bibinfo{year}{2009}).

\bibitem[{\citenamefont{Kamenetska et~al.}(2009)\citenamefont{Kamenetska,
  Koentopp, Whalley, Park, Steigerwald, Nuckolls, Hybertsen, and
  Venkataraman}}]{Kamenetska:2009}
\bibinfo{author}{\bibfnamefont{M.}~\bibnamefont{Kamenetska}},
  \bibinfo{author}{\bibfnamefont{M.}~\bibnamefont{Koentopp}},
  \bibinfo{author}{\bibfnamefont{A.~C.} \bibnamefont{Whalley}},
  \bibinfo{author}{\bibfnamefont{Y.~S.} \bibnamefont{Park}},
  \bibinfo{author}{\bibfnamefont{M.~L.} \bibnamefont{Steigerwald}},
  \bibinfo{author}{\bibfnamefont{C.}~\bibnamefont{Nuckolls}},
  \bibinfo{author}{\bibfnamefont{M.~S.} \bibnamefont{Hybertsen}},
  \bibnamefont{and}
  \bibinfo{author}{\bibfnamefont{L.}~\bibnamefont{Venkataraman}},
  \bibinfo{journal}{Phys. Rev. Lett.} \textbf{\bibinfo{volume}{102}},
  \bibinfo{pages}{126803} (\bibinfo{year}{2009}).

\bibitem[{\citenamefont{Vazquez et~al.}(2012)\citenamefont{Vazquez, Skouta,
  Schneebeli, Kamenetska, Breslow, Venkataraman, and Hybertsen}}]{Vazquez:2012}
\bibinfo{author}{\bibfnamefont{H.}~\bibnamefont{Vazquez}},
  \bibinfo{author}{\bibfnamefont{R.}~\bibnamefont{Skouta}},
  \bibinfo{author}{\bibfnamefont{S.}~\bibnamefont{Schneebeli}},
  \bibinfo{author}{\bibfnamefont{M.}~\bibnamefont{Kamenetska}},
  \bibinfo{author}{\bibfnamefont{R.}~\bibnamefont{Breslow}},
  \bibinfo{author}{\bibfnamefont{L.}~\bibnamefont{Venkataraman}},
  \bibnamefont{and} \bibinfo{author}{\bibfnamefont{M.~S.}
  \bibnamefont{Hybertsen}}, \bibinfo{journal}{Nat. Nanotechnol.}
  \textbf{\bibinfo{volume}{7}}, \bibinfo{pages}{663} (\bibinfo{year}{2012}).

\bibitem[{\citenamefont{L{\"o}rtscher et~al.}(2007)\citenamefont{L{\"o}rtscher,
  Weber, and Riel}}]{Lortscher:2007}
\bibinfo{author}{\bibfnamefont{E.}~\bibnamefont{L{\"o}rtscher}},
  \bibinfo{author}{\bibfnamefont{H.~B.} \bibnamefont{Weber}}, \bibnamefont{and}
  \bibinfo{author}{\bibfnamefont{H.}~\bibnamefont{Riel}},
  \bibinfo{journal}{Phys. Rev. Lett.} \textbf{\bibinfo{volume}{98}},
  \bibinfo{pages}{176807} (\bibinfo{year}{2007}).

\bibitem[{\citenamefont{Gu{\'e}don et~al.}(2012)\citenamefont{Gu{\'e}don,
  Valkenier, Markussen, Thygesen, Hummelen, and van~der Molen}}]{Guedon:2012}
\bibinfo{author}{\bibfnamefont{C.~M.} \bibnamefont{Gu{\'e}don}},
  \bibinfo{author}{\bibfnamefont{H.}~\bibnamefont{Valkenier}},
  \bibinfo{author}{\bibfnamefont{T.}~\bibnamefont{Markussen}},
  \bibinfo{author}{\bibfnamefont{K.~S.} \bibnamefont{Thygesen}},
  \bibinfo{author}{\bibfnamefont{J.~C.} \bibnamefont{Hummelen}},
  \bibnamefont{and} \bibinfo{author}{\bibfnamefont{S.~J.} \bibnamefont{van~der
  Molen}}, \bibinfo{journal}{Nat. Nanotechnol.} \textbf{\bibinfo{volume}{7}},
  \bibinfo{pages}{305} (\bibinfo{year}{2012}).

\bibitem[{\citenamefont{Darancet et~al.}(2012)\citenamefont{Darancet, Widawsky,
  Choi, Venkataraman, and Neaton}}]{Darancet:2012}
\bibinfo{author}{\bibfnamefont{P.}~\bibnamefont{Darancet}},
  \bibinfo{author}{\bibfnamefont{J.~R.} \bibnamefont{Widawsky}},
  \bibinfo{author}{\bibfnamefont{H.~J.} \bibnamefont{Choi}},
  \bibinfo{author}{\bibfnamefont{L.}~\bibnamefont{Venkataraman}},
  \bibnamefont{and} \bibinfo{author}{\bibfnamefont{J.~B.}
  \bibnamefont{Neaton}}, \bibinfo{journal}{Nano Lett.}
  \textbf{\bibinfo{volume}{12}}, \bibinfo{pages}{6250} (\bibinfo{year}{2012}).

\bibitem[{\citenamefont{Fournier et~al.}(2011)\citenamefont{Fournier, Wagner,
  Weiss, Temirov, and Tautz}}]{Fournier:2011}
\bibinfo{author}{\bibfnamefont{N.}~\bibnamefont{Fournier}},
  \bibinfo{author}{\bibfnamefont{C.}~\bibnamefont{Wagner}},
  \bibinfo{author}{\bibfnamefont{C.}~\bibnamefont{Weiss}},
  \bibinfo{author}{\bibfnamefont{R.}~\bibnamefont{Temirov}}, \bibnamefont{and}
  \bibinfo{author}{\bibfnamefont{F.~S.} \bibnamefont{Tautz}},
  \bibinfo{journal}{Phys. Rev. B} \textbf{\bibinfo{volume}{84}},
  \bibinfo{pages}{035435} (\bibinfo{year}{2011}).

\bibitem[{\citenamefont{Halbritter et~al.}(2010)\citenamefont{Halbritter, Makk,
  Mackowiak, Csonka, Wawrzyniak, and Martinek}}]{Halbritter:2010}
\bibinfo{author}{\bibfnamefont{A.}~\bibnamefont{Halbritter}},
  \bibinfo{author}{\bibfnamefont{P.}~\bibnamefont{Makk}},
  \bibinfo{author}{\bibfnamefont{S.}~\bibnamefont{Mackowiak}},
  \bibinfo{author}{\bibfnamefont{S.}~\bibnamefont{Csonka}},
  \bibinfo{author}{\bibfnamefont{M.}~\bibnamefont{Wawrzyniak}},
  \bibnamefont{and} \bibinfo{author}{\bibfnamefont{J.}~\bibnamefont{Martinek}},
  \bibinfo{journal}{Phys. Rev. Lett.} \textbf{\bibinfo{volume}{105}},
  \bibinfo{pages}{266805} (\bibinfo{year}{2010}).

\bibitem[{\citenamefont{Makk et~al.}(2012)\citenamefont{Makk, Tomaszewski,
  Martinek, Balogh, Csonka, Wawrzyniak, Frei, Venkataraman, and
  Halbritter}}]{Makk:2012}
\bibinfo{author}{\bibfnamefont{P.}~\bibnamefont{Makk}},
  \bibinfo{author}{\bibfnamefont{D.}~\bibnamefont{Tomaszewski}},
  \bibinfo{author}{\bibfnamefont{J.}~\bibnamefont{Martinek}},
  \bibinfo{author}{\bibfnamefont{Z.}~\bibnamefont{Balogh}},
  \bibinfo{author}{\bibfnamefont{S.}~\bibnamefont{Csonka}},
  \bibinfo{author}{\bibfnamefont{M.}~\bibnamefont{Wawrzyniak}},
  \bibinfo{author}{\bibfnamefont{M.}~\bibnamefont{Frei}},
  \bibinfo{author}{\bibfnamefont{L.}~\bibnamefont{Venkataraman}},
  \bibnamefont{and}
  \bibinfo{author}{\bibfnamefont{A.}~\bibnamefont{Halbritter}},
  \bibinfo{journal}{ACS Nano} \textbf{\bibinfo{volume}{6}},
  \bibinfo{pages}{3411} (\bibinfo{year}{2012}).

\bibitem[{\citenamefont{Aradhya et~al.}(2013)\citenamefont{Aradhya, Frei,
  Halbritter, and Venkataraman}}]{Aradhya:2013}
\bibinfo{author}{\bibfnamefont{S.~V.} \bibnamefont{Aradhya}},
  \bibinfo{author}{\bibfnamefont{M.}~\bibnamefont{Frei}},
  \bibinfo{author}{\bibfnamefont{A.}~\bibnamefont{Halbritter}},
  \bibnamefont{and}
  \bibinfo{author}{\bibfnamefont{L.}~\bibnamefont{Venkataraman}},
  \bibinfo{journal}{ACS Nano} \textbf{\bibinfo{volume}{7}},
  \bibinfo{pages}{3706} (\bibinfo{year}{2013}).

\bibitem[{\citenamefont{Dreher et~al.}(2005)\citenamefont{Dreher, Pauly,
  Heurich, Cuevas, Scheer, and Nielaba}}]{Dreher:2005}
\bibinfo{author}{\bibfnamefont{M.}~\bibnamefont{Dreher}},
  \bibinfo{author}{\bibfnamefont{F.}~\bibnamefont{Pauly}},
  \bibinfo{author}{\bibfnamefont{J.}~\bibnamefont{Heurich}},
  \bibinfo{author}{\bibfnamefont{J.~C.} \bibnamefont{Cuevas}},
  \bibinfo{author}{\bibfnamefont{E.}~\bibnamefont{Scheer}}, \bibnamefont{and}
  \bibinfo{author}{\bibfnamefont{P.}~\bibnamefont{Nielaba}},
  \bibinfo{journal}{Phys. Rev. B} \textbf{\bibinfo{volume}{72}},
  \bibinfo{pages}{075435} (\bibinfo{year}{2005}).

\bibitem[{\citenamefont{Pauly et~al.}(2006)\citenamefont{Pauly, Dreher, Viljas,
  H\"afner, Cuevas, and Nielaba}}]{Pauly:2006}
\bibinfo{author}{\bibfnamefont{F.}~\bibnamefont{Pauly}},
  \bibinfo{author}{\bibfnamefont{M.}~\bibnamefont{Dreher}},
  \bibinfo{author}{\bibfnamefont{J.~K.} \bibnamefont{Viljas}},
  \bibinfo{author}{\bibfnamefont{M.}~\bibnamefont{H\"afner}},
  \bibinfo{author}{\bibfnamefont{J.~C.} \bibnamefont{Cuevas}},
  \bibnamefont{and} \bibinfo{author}{\bibfnamefont{P.}~\bibnamefont{Nielaba}},
  \bibinfo{journal}{Phys. Rev. B} \textbf{\bibinfo{volume}{74}},
  \bibinfo{pages}{235106} (\bibinfo{year}{2006}).

\bibitem[{\citenamefont{Makk et~al.}(2011)\citenamefont{Makk, Visontai,
  Oroszl\'any, Manrique, Csonka, Cserti, Lambert, and Halbritter}}]{Makk:2011}
\bibinfo{author}{\bibfnamefont{P.}~\bibnamefont{Makk}},
  \bibinfo{author}{\bibfnamefont{D.}~\bibnamefont{Visontai}},
  \bibinfo{author}{\bibfnamefont{L.}~\bibnamefont{Oroszl\'any}},
  \bibinfo{author}{\bibfnamefont{D.~Z.} \bibnamefont{Manrique}},
  \bibinfo{author}{\bibfnamefont{S.}~\bibnamefont{Csonka}},
  \bibinfo{author}{\bibfnamefont{J.}~\bibnamefont{Cserti}},
  \bibinfo{author}{\bibfnamefont{C.}~\bibnamefont{Lambert}}, \bibnamefont{and}
  \bibinfo{author}{\bibfnamefont{A.}~\bibnamefont{Halbritter}},
  \bibinfo{journal}{Phys. Rev. Lett.} \textbf{\bibinfo{volume}{107}},
  \bibinfo{pages}{276801} (\bibinfo{year}{2011}).

\bibitem[{\citenamefont{Schirm et~al.}(2013)\citenamefont{Schirm, Matt, Pauly,
  Cuevas, Nielaba, and Scheer}}]{Schirm:2013}
\bibinfo{author}{\bibfnamefont{C.}~\bibnamefont{Schirm}},
  \bibinfo{author}{\bibfnamefont{M.}~\bibnamefont{Matt}},
  \bibinfo{author}{\bibfnamefont{F.}~\bibnamefont{Pauly}},
  \bibinfo{author}{\bibfnamefont{J.~C.} \bibnamefont{Cuevas}},
  \bibinfo{author}{\bibfnamefont{P.}~\bibnamefont{Nielaba}}, \bibnamefont{and}
  \bibinfo{author}{\bibfnamefont{E.}~\bibnamefont{Scheer}},
  \bibinfo{journal}{Nat. Nanotechnol.} \textbf{\bibinfo{volume}{8}},
  \bibinfo{pages}{645} (\bibinfo{year}{2013}).

\bibitem[{\citenamefont{Wheeler et~al.}(2010)\citenamefont{Wheeler, Russom,
  Evans, King, and Natelson}}]{Wheeler:2010}
\bibinfo{author}{\bibfnamefont{P.}~\bibnamefont{Wheeler}},
  \bibinfo{author}{\bibfnamefont{J.}~\bibnamefont{Russom}},
  \bibinfo{author}{\bibfnamefont{K.}~\bibnamefont{Evans}},
  \bibinfo{author}{\bibfnamefont{N.}~\bibnamefont{King}}, \bibnamefont{and}
  \bibinfo{author}{\bibfnamefont{D.}~\bibnamefont{Natelson}},
  \bibinfo{journal}{Nano Lett.} \textbf{\bibinfo{volume}{10}},
  \bibinfo{pages}{1287} (\bibinfo{year}{2010}).

\bibitem[{\citenamefont{Chen et~al.}(2012)\citenamefont{Chen, Wheeler, and
  Natelson}}]{Chen:2012}
\bibinfo{author}{\bibfnamefont{R.}~\bibnamefont{Chen}},
  \bibinfo{author}{\bibfnamefont{P.~J.} \bibnamefont{Wheeler}},
  \bibnamefont{and} \bibinfo{author}{\bibfnamefont{D.}~\bibnamefont{Natelson}},
  \bibinfo{journal}{Phys. Rev. B} \textbf{\bibinfo{volume}{85}},
  \bibinfo{pages}{235455} (\bibinfo{year}{2012}).

\bibitem[{\citenamefont{Chen et~al.}(2014)\citenamefont{Chen, Wheeler,
  Di~Ventra, and Natelson}}]{Chen:2014}
\bibinfo{author}{\bibfnamefont{R.}~\bibnamefont{Chen}},
  \bibinfo{author}{\bibfnamefont{P.~J.} \bibnamefont{Wheeler}},
  \bibinfo{author}{\bibfnamefont{M.}~\bibnamefont{Di~Ventra}},
  \bibnamefont{and} \bibinfo{author}{\bibfnamefont{D.}~\bibnamefont{Natelson}},
  \bibinfo{journal}{Sci. Rep.} \textbf{\bibinfo{volume}{4}},
  \bibinfo{pages}{4221} (\bibinfo{year}{2014}).

\bibitem[{\citenamefont{Schottky}(1918)}]{Schottky:1918}
\bibinfo{author}{\bibfnamefont{W.}~\bibnamefont{Schottky}},
  \bibinfo{journal}{Ann. Phys.} \textbf{\bibinfo{volume}{57}},
  \bibinfo{pages}{541} (\bibinfo{year}{1918}).

\bibitem[{\citenamefont{Blanter and B{\"u}ttiker}(2000)}]{Blanter:2000}
\bibinfo{author}{\bibfnamefont{Y.}~\bibnamefont{Blanter}} \bibnamefont{and}
  \bibinfo{author}{\bibfnamefont{M.}~\bibnamefont{B{\"u}ttiker}},
  \bibinfo{journal}{Phys. Rep.} \textbf{\bibinfo{volume}{336}},
  \bibinfo{pages}{1 } (\bibinfo{year}{2000}).

\bibitem[{\citenamefont{Johnson}(1928)}]{Johnson:1928}
\bibinfo{author}{\bibfnamefont{J.~B.} \bibnamefont{Johnson}},
  \bibinfo{journal}{Phys. Rev.} \textbf{\bibinfo{volume}{32}},
  \bibinfo{pages}{97} (\bibinfo{year}{1928}).

\bibitem[{\citenamefont{Nyquist}(1928)}]{Nyquist:1928}
\bibinfo{author}{\bibfnamefont{H.}~\bibnamefont{Nyquist}},
  \bibinfo{journal}{Phys. Rev.} \textbf{\bibinfo{volume}{32}},
  \bibinfo{pages}{110} (\bibinfo{year}{1928}).

\bibitem[{\citenamefont{de~Picciotto et~al.}(1997)\citenamefont{de~Picciotto,
  Reznikov, Heiblum, Umansky, Bunin, and Mahalu}}]{dePicciotto:1997}
\bibinfo{author}{\bibfnamefont{R.}~\bibnamefont{de~Picciotto}},
  \bibinfo{author}{\bibfnamefont{M.}~\bibnamefont{Reznikov}},
  \bibinfo{author}{\bibfnamefont{M.}~\bibnamefont{Heiblum}},
  \bibinfo{author}{\bibfnamefont{V.}~\bibnamefont{Umansky}},
  \bibinfo{author}{\bibfnamefont{G.}~\bibnamefont{Bunin}}, \bibnamefont{and}
  \bibinfo{author}{\bibfnamefont{D.}~\bibnamefont{Mahalu}},
  \bibinfo{journal}{Nature} \textbf{\bibinfo{volume}{389}},
  \bibinfo{pages}{162} (\bibinfo{year}{1997}).

\bibitem[{\citenamefont{Saminadayar et~al.}(1997)\citenamefont{Saminadayar,
  Glattli, Jin, and Etienne}}]{Saminadayar:1997}
\bibinfo{author}{\bibfnamefont{L.}~\bibnamefont{Saminadayar}},
  \bibinfo{author}{\bibfnamefont{D.~C.} \bibnamefont{Glattli}},
  \bibinfo{author}{\bibfnamefont{Y.}~\bibnamefont{Jin}}, \bibnamefont{and}
  \bibinfo{author}{\bibfnamefont{B.}~\bibnamefont{Etienne}},
  \bibinfo{journal}{Phys. Rev. Lett.} \textbf{\bibinfo{volume}{79}},
  \bibinfo{pages}{2526} (\bibinfo{year}{1997}).

\bibitem[{\citenamefont{Datta}(1995)}]{Datta:1995}
\bibinfo{author}{\bibfnamefont{S.}~\bibnamefont{Datta}},
  \emph{\bibinfo{title}{Electronic Transport in Mesoscopic Systems}}
  (\bibinfo{publisher}{Cambridge University Press},
  \bibinfo{address}{Cambridge}, \bibinfo{year}{1995}).

\bibitem[{\citenamefont{Di~Ventra}(2008)}]{DiVentra:2008}
\bibinfo{author}{\bibfnamefont{M.}~\bibnamefont{Di~Ventra}},
  \emph{\bibinfo{title}{Electrical Transport in Nanoscale Systems}}
  (\bibinfo{publisher}{Cambridge University Press},
  \bibinfo{address}{Cambridge}, \bibinfo{year}{2008}).

\bibitem[{\citenamefont{Cuevas and Scheer}(2010)}]{Cuevas:2010}
\bibinfo{author}{\bibfnamefont{J.~C.} \bibnamefont{Cuevas}} \bibnamefont{and}
  \bibinfo{author}{\bibfnamefont{E.}~\bibnamefont{Scheer}},
  \emph{\bibinfo{title}{Molecular Electronics: An Introduction to Theory and
  Experiment}} (\bibinfo{publisher}{World Scientific},
  \bibinfo{address}{Singapore}, \bibinfo{year}{2010}).

\bibitem[{\citenamefont{Khlus}(1987)}]{Khlus:1987}
\bibinfo{author}{\bibfnamefont{V.}~\bibnamefont{Khlus}}, \bibinfo{journal}{Sov.
  Phys. JETP} \textbf{\bibinfo{volume}{66}}, \bibinfo{pages}{1243}
  (\bibinfo{year}{1987}).

\bibitem[{\citenamefont{Lesovik}(1989)}]{Lesovik:1989}
\bibinfo{author}{\bibfnamefont{G.~B.} \bibnamefont{Lesovik}},
  \bibinfo{journal}{Sov. Phys. JETP Lett.} \textbf{\bibinfo{volume}{49}},
  \bibinfo{pages}{592} (\bibinfo{year}{1989}).

\bibitem[{\citenamefont{Landauer and Martin}(1991)}]{Landauer:1991}
\bibinfo{author}{\bibfnamefont{R.}~\bibnamefont{Landauer}} \bibnamefont{and}
  \bibinfo{author}{\bibfnamefont{T.}~\bibnamefont{Martin}},
  \bibinfo{journal}{Physica B} \textbf{\bibinfo{volume}{175}},
  \bibinfo{pages}{167 } (\bibinfo{year}{1991}).

\bibitem[{\citenamefont{Martin and Landauer}(1992)}]{Martin:1992}
\bibinfo{author}{\bibfnamefont{T.}~\bibnamefont{Martin}} \bibnamefont{and}
  \bibinfo{author}{\bibfnamefont{R.}~\bibnamefont{Landauer}},
  \bibinfo{journal}{Phys. Rev. B} \textbf{\bibinfo{volume}{45}},
  \bibinfo{pages}{1742} (\bibinfo{year}{1992}).

\bibitem[{\citenamefont{B{\"u}ttiker}(1992)}]{Buttiker:1992}
\bibinfo{author}{\bibfnamefont{M.}~\bibnamefont{B{\"u}ttiker}},
  \bibinfo{journal}{Phys. Rev. B} \textbf{\bibinfo{volume}{46}},
  \bibinfo{pages}{12485} (\bibinfo{year}{1992}).

\bibitem[{\citenamefont{Ludoph and van Ruitenbeek}(2000)}]{Ludoph:2000}
\bibinfo{author}{\bibfnamefont{B.}~\bibnamefont{Ludoph}} \bibnamefont{and}
  \bibinfo{author}{\bibfnamefont{J.~M.} \bibnamefont{van Ruitenbeek}},
  \bibinfo{journal}{Phys. Rev. B} \textbf{\bibinfo{volume}{61}},
  \bibinfo{pages}{2273} (\bibinfo{year}{2000}).

\bibitem[{\citenamefont{Ludoph et~al.}(1999)\citenamefont{Ludoph, Devoret,
  Esteve, Urbina, and van Ruitenbeek}}]{Ludoph:1999}
\bibinfo{author}{\bibfnamefont{B.}~\bibnamefont{Ludoph}},
  \bibinfo{author}{\bibfnamefont{M.~H.} \bibnamefont{Devoret}},
  \bibinfo{author}{\bibfnamefont{D.}~\bibnamefont{Esteve}},
  \bibinfo{author}{\bibfnamefont{C.}~\bibnamefont{Urbina}}, \bibnamefont{and}
  \bibinfo{author}{\bibfnamefont{J.~M.} \bibnamefont{van Ruitenbeek}},
  \bibinfo{journal}{Phys. Rev. Lett.} \textbf{\bibinfo{volume}{82}},
  \bibinfo{pages}{1530} (\bibinfo{year}{1999}).

\bibitem[{\citenamefont{van~den Brom and van
  Ruitenbeek}(1999)}]{vandenBrom:1999}
\bibinfo{author}{\bibfnamefont{H.~E.} \bibnamefont{van~den Brom}}
  \bibnamefont{and} \bibinfo{author}{\bibfnamefont{J.~M.} \bibnamefont{van
  Ruitenbeek}}, \bibinfo{journal}{Phys. Rev. Lett.}
  \textbf{\bibinfo{volume}{82}}, \bibinfo{pages}{1526} (\bibinfo{year}{1999}).

\bibitem[{\citenamefont{Pauly et~al.}(2011)\citenamefont{Pauly, Viljas,
  B\"urkle, Dreher, Nielaba, and Cuevas}}]{Pauly:2011}
\bibinfo{author}{\bibfnamefont{F.}~\bibnamefont{Pauly}},
  \bibinfo{author}{\bibfnamefont{J.~K.} \bibnamefont{Viljas}},
  \bibinfo{author}{\bibfnamefont{M.}~\bibnamefont{B\"urkle}},
  \bibinfo{author}{\bibfnamefont{M.}~\bibnamefont{Dreher}},
  \bibinfo{author}{\bibfnamefont{P.}~\bibnamefont{Nielaba}}, \bibnamefont{and}
  \bibinfo{author}{\bibfnamefont{J.~C.} \bibnamefont{Cuevas}},
  \bibinfo{journal}{Phys. Rev. B} \textbf{\bibinfo{volume}{84}},
  \bibinfo{pages}{195420} (\bibinfo{year}{2011}).

\bibitem[{\citenamefont{Plimpton}(1995)}]{Plimpton:1995}
\bibinfo{author}{\bibfnamefont{S.}~\bibnamefont{Plimpton}},
  \bibinfo{journal}{J. Comput. Phys.} \textbf{\bibinfo{volume}{117}},
  \bibinfo{pages}{1 } (\bibinfo{year}{1995}).

\bibitem[{LAM()}]{LAMMPS}
\emph{\bibinfo{title}{http://lammps.sandia.gov}}.

\bibitem[{\citenamefont{Sheng et~al.}(2011)\citenamefont{Sheng, Kramer, Cadien,
  Fujita, and Chen}}]{Sheng:2011}
\bibinfo{author}{\bibfnamefont{H.~W.} \bibnamefont{Sheng}},
  \bibinfo{author}{\bibfnamefont{M.~J.} \bibnamefont{Kramer}},
  \bibinfo{author}{\bibfnamefont{A.}~\bibnamefont{Cadien}},
  \bibinfo{author}{\bibfnamefont{T.}~\bibnamefont{Fujita}}, \bibnamefont{and}
  \bibinfo{author}{\bibfnamefont{M.~W.} \bibnamefont{Chen}},
  \bibinfo{journal}{Phys. Rev. B} \textbf{\bibinfo{volume}{83}},
  \bibinfo{pages}{134118} (\bibinfo{year}{2011}).

\bibitem[{\citenamefont{Frenkel and Smit}(2004)}]{Frenkel:2004}
\bibinfo{author}{\bibfnamefont{D.}~\bibnamefont{Frenkel}} \bibnamefont{and}
  \bibinfo{author}{\bibfnamefont{B.}~\bibnamefont{Smit}},
  \emph{\bibinfo{title}{Understanding molecular simulation}}
  (\bibinfo{publisher}{Academic Press}, \bibinfo{address}{San Diego},
  \bibinfo{year}{2004}).

\bibitem[{\citenamefont{Pauly et~al.}(2008)\citenamefont{Pauly, Viljas, Huniar,
  H\"afner, Wohlthat, B\"urkle, Cuevas, and Sch\"on}}]{Pauly:2008}
\bibinfo{author}{\bibfnamefont{F.}~\bibnamefont{Pauly}},
  \bibinfo{author}{\bibfnamefont{J.}~\bibnamefont{Viljas}},
  \bibinfo{author}{\bibfnamefont{U.}~\bibnamefont{Huniar}},
  \bibinfo{author}{\bibfnamefont{M.}~\bibnamefont{H\"afner}},
  \bibinfo{author}{\bibfnamefont{S.}~\bibnamefont{Wohlthat}},
  \bibinfo{author}{\bibfnamefont{M.}~\bibnamefont{B\"urkle}},
  \bibinfo{author}{\bibfnamefont{J.~C.} \bibnamefont{Cuevas}},
  \bibnamefont{and} \bibinfo{author}{\bibfnamefont{G.}~\bibnamefont{Sch\"on}},
  \bibinfo{journal}{New J. Phys.} \textbf{\bibinfo{volume}{10}},
  \bibinfo{pages}{125019} (\bibinfo{year}{2008}).

\bibitem[{\citenamefont{Mehl and Papaconstantopoulos}(1998)}]{Mehl:1998}
\bibinfo{author}{\bibfnamefont{M.~J.} \bibnamefont{Mehl}} \bibnamefont{and}
  \bibinfo{author}{\bibfnamefont{D.~A.} \bibnamefont{Papaconstantopoulos}},
  \emph{\bibinfo{title}{Computational Materials Science (edited by C. Fong)}}
  (\bibinfo{publisher}{World Scientific}, \bibinfo{address}{Singapore},
  \bibinfo{year}{1998}).

\bibitem[{\citenamefont{Guinea et~al.}(1983)\citenamefont{Guinea, Tejedor,
  Flores, and Louis}}]{Guinea:1983}
\bibinfo{author}{\bibfnamefont{F.}~\bibnamefont{Guinea}},
  \bibinfo{author}{\bibfnamefont{C.}~\bibnamefont{Tejedor}},
  \bibinfo{author}{\bibfnamefont{F.}~\bibnamefont{Flores}}, \bibnamefont{and}
  \bibinfo{author}{\bibfnamefont{E.}~\bibnamefont{Louis}},
  \bibinfo{journal}{Phys. Rev. B} \textbf{\bibinfo{volume}{28}},
  \bibinfo{pages}{4397} (\bibinfo{year}{1983}).

\bibitem[{\citenamefont{Brandbyge et~al.}(1999)\citenamefont{Brandbyge,
  Kobayashi, and Tsukada}}]{Brandbyge:1999}
\bibinfo{author}{\bibfnamefont{M.}~\bibnamefont{Brandbyge}},
  \bibinfo{author}{\bibfnamefont{N.}~\bibnamefont{Kobayashi}},
  \bibnamefont{and} \bibinfo{author}{\bibfnamefont{M.}~\bibnamefont{Tsukada}},
  \bibinfo{journal}{Phys. Rev. B} \textbf{\bibinfo{volume}{60}},
  \bibinfo{pages}{17064} (\bibinfo{year}{1999}).

\bibitem[{\citenamefont{Ohnishi et~al.}(1998)\citenamefont{Ohnishi, Kondo, and
  Takayanagi}}]{Ohnishi:1998}
\bibinfo{author}{\bibfnamefont{H.}~\bibnamefont{Ohnishi}},
  \bibinfo{author}{\bibfnamefont{Y.}~\bibnamefont{Kondo}}, \bibnamefont{and}
  \bibinfo{author}{\bibfnamefont{K.}~\bibnamefont{Takayanagi}},
  \bibinfo{journal}{Nature} \textbf{\bibinfo{volume}{395}},
  \bibinfo{pages}{780} (\bibinfo{year}{1998}).

\bibitem[{\citenamefont{Yanson et~al.}(1998)\citenamefont{Yanson, Bollinger,
  van~den Brom, Agra\"it, and van Ruitenbeek}}]{Yanson:1998}
\bibinfo{author}{\bibfnamefont{A.~I.} \bibnamefont{Yanson}},
  \bibinfo{author}{\bibfnamefont{G.~R.} \bibnamefont{Bollinger}},
  \bibinfo{author}{\bibfnamefont{H.~E.} \bibnamefont{van~den Brom}},
  \bibinfo{author}{\bibfnamefont{N.}~\bibnamefont{Agra\"it}}, \bibnamefont{and}
  \bibinfo{author}{\bibfnamefont{J.~M.} \bibnamefont{van Ruitenbeek}},
  \bibinfo{journal}{Nature} \textbf{\bibinfo{volume}{395}},
  \bibinfo{pages}{783} (\bibinfo{year}{1998}).

\bibitem[{\citenamefont{Vardimon and Tal}(2014)}]{Vardimon:2013a}
\bibinfo{author}{\bibfnamefont{R.}~\bibnamefont{Vardimon}} \bibnamefont{and}
  \bibinfo{author}{\bibfnamefont{O.}~\bibnamefont{Tal}},
  \bibinfo{howpublished}{private communication} (\bibinfo{year}{2014}).

\bibitem[{\citenamefont{Gai et~al.}(1996)\citenamefont{Gai, He, Yu, and
  Yang}}]{Gai:1996}
\bibinfo{author}{\bibfnamefont{Z.}~\bibnamefont{Gai}},
  \bibinfo{author}{\bibfnamefont{Y.}~\bibnamefont{He}},
  \bibinfo{author}{\bibfnamefont{H.}~\bibnamefont{Yu}}, \bibnamefont{and}
  \bibinfo{author}{\bibfnamefont{W.~S.} \bibnamefont{Yang}},
  \bibinfo{journal}{Phys. Rev. B} \textbf{\bibinfo{volume}{53}},
  \bibinfo{pages}{1042} (\bibinfo{year}{1996}).

\bibitem[{\citenamefont{Costa-Kr\"amer}(1997)}]{CostaKramer:1997}
\bibinfo{author}{\bibfnamefont{J.~L.} \bibnamefont{Costa-Kr\"amer}},
  \bibinfo{journal}{Phys. Rev. B} \textbf{\bibinfo{volume}{55}},
  \bibinfo{pages}{R4875} (\bibinfo{year}{1997}).

\bibitem[{\citenamefont{Csonka et~al.}(2003)\citenamefont{Csonka, Halbritter,
  Mih\'aly, Jurdik, Shklyarevskii, Speller, and van Kempen}}]{Csonka:2003}
\bibinfo{author}{\bibfnamefont{S.}~\bibnamefont{Csonka}},
  \bibinfo{author}{\bibfnamefont{A.}~\bibnamefont{Halbritter}},
  \bibinfo{author}{\bibfnamefont{G.}~\bibnamefont{Mih\'aly}},
  \bibinfo{author}{\bibfnamefont{E.}~\bibnamefont{Jurdik}},
  \bibinfo{author}{\bibfnamefont{O.~I.} \bibnamefont{Shklyarevskii}},
  \bibinfo{author}{\bibfnamefont{S.}~\bibnamefont{Speller}}, \bibnamefont{and}
  \bibinfo{author}{\bibfnamefont{H.}~\bibnamefont{van Kempen}},
  \bibinfo{journal}{Phys. Rev. Lett.} \textbf{\bibinfo{volume}{90}},
  \bibinfo{pages}{116803} (\bibinfo{year}{2003}).

\bibitem[{\citenamefont{Barnett et~al.}(2004)\citenamefont{Barnett,
  H{\"a}kkinen, Scherbakov, and Landman}}]{Barnett:2004}
\bibinfo{author}{\bibfnamefont{R.~N.} \bibnamefont{Barnett}},
  \bibinfo{author}{\bibfnamefont{H.}~\bibnamefont{H{\"a}kkinen}},
  \bibinfo{author}{\bibfnamefont{A.~G.} \bibnamefont{Scherbakov}},
  \bibnamefont{and} \bibinfo{author}{\bibfnamefont{U.}~\bibnamefont{Landman}},
  \bibinfo{journal}{Nano Letters} \textbf{\bibinfo{volume}{4}},
  \bibinfo{pages}{1845} (\bibinfo{year}{2004}).

\bibitem[{\citenamefont{Thijssen et~al.}(2008)\citenamefont{Thijssen, Strange,
  aan~de Brugh, and van Ruitenbeek}}]{Thijssen:2008}
\bibinfo{author}{\bibfnamefont{W.~H.~A.} \bibnamefont{Thijssen}},
  \bibinfo{author}{\bibfnamefont{M.}~\bibnamefont{Strange}},
  \bibinfo{author}{\bibfnamefont{J.~M.~J.} \bibnamefont{aan~de Brugh}},
  \bibnamefont{and} \bibinfo{author}{\bibfnamefont{J.~M.} \bibnamefont{van
  Ruitenbeek}}, \bibinfo{journal}{New J. of Phys.}
  \textbf{\bibinfo{volume}{10}}, \bibinfo{pages}{033005}
  (\bibinfo{year}{2008}).

\end{thebibliography}

\end{document}